\documentclass[preprint,12pt]{elsarticle}

\usepackage{amssymb}
\usepackage{amsmath}
\usepackage{array}
\usepackage{subcaption}%
\usepackage{graphicx}%
\usepackage{multirow}%
\usepackage{amsmath,amssymb,amsfonts}%
\usepackage{amsthm}%
\usepackage{mathrsfs}%
\usepackage[title]{appendix}%
\usepackage{xcolor}%
\usepackage{textcomp}%
\usepackage{manyfoot}%
\usepackage{booktabs}%
\usepackage{algorithm}%
\usepackage{algorithmicx}%
\usepackage{algpseudocode}%
\usepackage{listings}%
\newcolumntype{P}[1]{>{\centering\arraybackslash}p{#1}}

\journal{}

\begin{document}

\begin{frontmatter}



\title{An Empirical Evaluation of AI-Powered Non-Player Characters’ Perceived Realism and Performance in Virtual Reality Environments}


\author{Mikko Korkiakoski} 
\author{Saeid Sheikhi}
\author{Jesper Nyman}
\author{Jussi Saariniemi}
\author{Kalle Tapio}
\author{Panos Kostakos}

\affiliation{organization={Center for Ubiquitous Computing, University of Oulu},
            addressline={Erkki Koiso-Kanttilan katu 3}, 
            city={Oulu},
            postcode={90014},            
            country={Finland}}

\begin{abstract}
Advancements in artificial intelligence (AI) have significantly enhanced the realism and interactivity of non-player characters (NPCs) in virtual reality (VR), creating more engaging and believable user experiences. This paper evaluates AI-driven NPCs within a VR interrogation simulator, focusing on their perceived realism, usability, and system performance. The simulator features two AI-powered NPCs, a suspect, and a partner, using GPT-4 Turbo to engage participants in a scenario to determine the suspect's guilt or innocence. A user study with 18 participants assessed the system using the System Usability Scale (SUS), Game Experience Questionnaire (GEQ), and a Virtual Agent Believability Questionnaire, alongside latency measurements for speech-to-text (STT), text-to-speech (TTS), OpenAI GPT-4 Turbo, and overall (cycle) latency. Results showed an average cycle latency of 7 seconds, influenced by the increasing conversational context. Believability scored 6.67 out of 10, with high ratings in behavior, social relationships, and intelligence but moderate scores in emotion and personality. The system achieved a SUS score of 79.44, indicating good usability. These findings demonstrate the potential of large language models to improve NPC realism and interaction in VR while highlighting challenges in reducing system latency and enhancing emotional depth. This research contributes to the development of more sophisticated AI-driven NPCs, revealing the need for performance optimization to achieve increasingly immersive virtual experiences.
\end{abstract}


\begin{highlights}
\item Integrated GPT-4 Turbo for realistic AI NPCs in VR
\item Developed a VR interrogation simulator with AI-controlled characters
\item Conducted an empirical study on NPC believability and usability
\item Identified latency challenges impacting real-time interactions
\end{highlights}

\begin{keyword}
artificial intelligence \sep large language model \sep virtual reality \sep non-player character


\end{keyword}

\end{frontmatter}



\section{Introduction}\label{sec1}

The integration of artificial intelligence (AI) into the development of non-player characters (NPCs) has significantly transformed the gaming landscape, particularly within virtual reality (VR) environments. The evolution of NPCs from simple scripted entities to complex, AI-driven characters capable of dynamic interactions has been a focal point of research in game AI. Traditional decision-making methods for NPCs, such as state machines and behavior trees, are increasingly supplemented or replaced by more sophisticated AI techniques, including machine learning and procedural content generation (PCG) (\cite{lapeyrade2023non,karaca2023ai,filipovic2023role}. These advancements enable NPCs to exhibit more human-like behaviors, enhancing the realism and immersion of gaming experiences.
The perceived realism of NPCs in VR is crucial for player engagement and satisfaction. As NPCs become more realistic, players expect them to respond in believable and contextually appropriate manners. Research indicates that NPCs can be designed to possess emotional and cognitive intelligence, allowing them to react dynamically to player actions \cite{karaca2023ai,zeng2023review}. The incorporation of AI further enhances this capability by enabling NPCs to generate contextually relevant and nuanced dialogues. This not only enriches the gameplay experience but also fosters a deeper emotional connection between players and the game world. For instance, NPCs that provide feedback or exhibit spectator-like behaviors in VR settings have been shown to influence player performance and overall experience \cite{xu2023exploring,guo2023s}. Such interactions are essential for creating immersive environments where players feel a sense of agency and realism.\\
Furthermore, the application of AI in NPC development extends beyond behavior simulation to encompass complex social interactions within games. Designing NPCs for VR-based social skill training, particularly for populations such as autistic children, highlights the importance of social embodiment in facilitating meaningful interactions \cite{moon2018reviews}. The use of AI models in these contexts allows for more adaptive and personalized interactions, enhancing the effectiveness of educational and therapeutic interventions. This underscores the role of NPCs not just as game elements but as integral components of various applications, demonstrating their versatility and potential impact across different contexts. The technical foundations of AI-driven NPCs involve sophisticated algorithms that facilitate adaptive behavior and learning from player interactions. The use of finite state machines (FSM) and machine learning models enables NPCs to make diverse decisions and adapt to changing environments, thereby enhancing their realism \cite{riyan2023implementation}. Moreover, ongoing research into AI-based NPCs emphasizes the need for efficient algorithms that can operate within the constraints of VR systems, ensuring that NPCs can perform complex tasks without compromising overall game performance \cite{ribeiro2023virtual,zheng2024memoryrepository}.\\
In recent years, AI technologies have seen significant advancements, particularly in large language models (LLMs), which can greatly enhance the capabilities of AI-driven NPCs. These models can improve NPC performance in handling complex tasks and facilitate adaptive behavior, enabling more dynamic learning from player interactions.  By leveraging LLMs, NPCs can process natural language inputs and generate responses that are contextually appropriate and linguistically coherent. This advancement enables NPCs to make diverse decisions and adapt to changing environments, thereby enhancing their realism. Moreover, integrating LLMs into VR systems presents challenges related to computational efficiency and performance constraints. Ongoing research emphasizes the need for efficient algorithms that can operate within these constraints, ensuring that NPCs can perform complex tasks without compromising overall game performance. Therefore, this paper presents an empirical evaluation of AI-powered NPCs' perceived realism and performance in VR environments, focusing on the integration of LLMs in development. Our study uses a complex relationship between advanced AI technologies and player experience. As AI continues to evolve, the potential for NPCs to provide realistic, engaging, and emotionally resonant interactions will likely expand, further blurring the lines between virtual and real-world experiences.

\section{Related works}\label{sec2}

The literature on AI-powered NPCs in virtual reality environments reveals that perceived realism and performance are critical factors in user experience, heavily influenced by technological advancements and user interactions. Research has consistently highlighted the importance of NPC design in creating immersive environments that cognitively and emotionally engage users.
AI-powered NPCs enhance user engagement by mimicking human-like behavior, which in turn increases the level of immersion. Xu et al. (2023) \cite{xu2023exploring} demonstrate the impact of NPC spectators in VR exergames, showing that their presence can influence player performance and overall experience. Similarly, Guo (2023) \cite{guo2023s} found that the familiarity of NPC audiences directly affects player exertion and enjoyment, highlighting the role of NPC appearance, voice, and behavior in creating a relatable and immersive experience. These elements are crucial for enhancing the perceived realism of VR environments, making the NPCs feel more natural and integrated into the virtual world.
Technological advancements in AI have played a significant role in improving NPC realism. The complexity of AI algorithms used to simulate NPC actions significantly affects user perception. For instance, Santoso (2023) \cite{ong2022dragonfly} introduced the Dragonfly Algorithm, which enhances NPC movement in crowded environments, contributing to greater realism and user engagement. Such advancements enable NPCs to behave in more natural, human-like ways, making the VR environment feel more dynamic and interactive.
Furthermore, NPCs are not only passive entities; they actively shape gameplay and user performance. Strojny et al. (2020) \cite{strojny2020moderators} explored the social facilitation effects of virtual agents, showing that the realism of NPCs can significantly influence user behavior. Kim et al. (2020) \cite{kim2020navigating} contributed to this by developing a method that allows NPCs to navigate based on user actions, creating a more responsive environment. These studies highlight that NPCs can adapt to player behavior, enhancing the sense of presence and improving the overall quality of the VR experience.
\\\\
In addition to technical sophistication, the emotional and psychological impact of interacting with NPCs in VR is crucial for user immersion. Breves (2018) \cite{breves2020reducing} explored how diverse NPC representations can reduce biases and increase empathy among players, suggesting that NPC design should not only focus on realism but also consider the social implications of these interactions. Emotional connections between users and NPCs can strengthen the overall sense of immersion and engagement, making the virtual environment feel more authentic and meaningful.
While these advancements demonstrate considerable progress, challenges remain. Oumaima et al. (2023) \cite{oumaima2023application} discusses the potential of AI to improve VR interactivity and fidelity but also stresses the need for ethical considerations, particularly regarding how AI influences user behavior and perceptions. Pashentsev (2023) \cite{pashentsev2023metaverses} highlights the ethical difficulties of AI use in VR, including the risk of bias and user manipulation. These concerns suggest that while technological advancements are essential, they must be balanced with considerations of responsible AI use in virtual environments.\\\\
In line with these findings, this research investigates the integration of generative AI (GenAI) with VR to create more realistic NPCs. Our study focuses on developing NPCs capable of dynamic, context-aware interactions driven by state-of-the-art AI technologies such as speech-to-text, text-to-speech, and generative pre-trained transformer (GPT) models. Our objective is to enhance the realism and immersion of NPCs within VR environments, thereby providing users with a more interactive and engaging experience. This work contributes to the ongoing exploration of AI's potential to transform the virtual reality landscape by creating NPCs that are not only realistic but also responsive to player actions, further improving the interactive nature of VR environments.

\section{The interrogation simulator}

\begin{figure*}[h!]
    \centering
    \begin{subfigure}[t]{0.5\linewidth}
        \centering
        \includegraphics[width=1\linewidth]{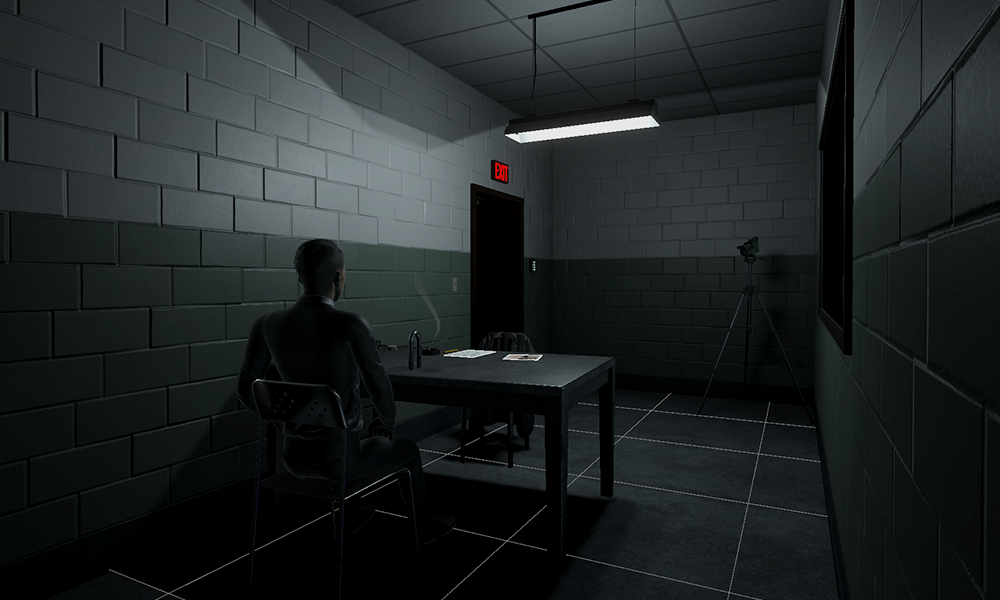}
        \caption{The interrogation room and suspect NPC}
    \end{subfigure}%
    ~ 
    \begin{subfigure}[t]{0.5\linewidth}
        \centering
        \includegraphics[width=1\linewidth]{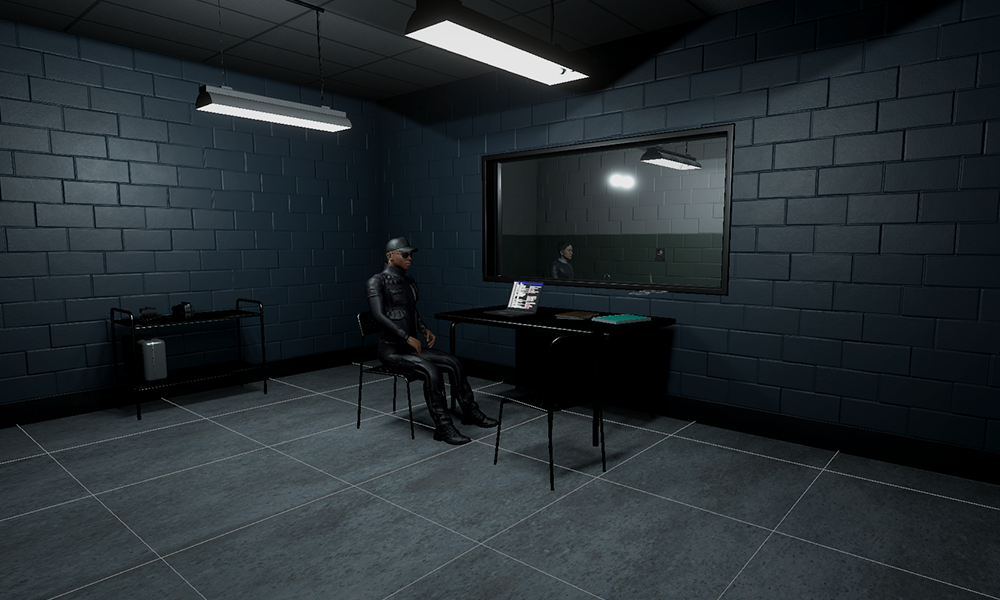}
        \caption{The observation room and partner NPC}
    \end{subfigure}
    \centering
    \begin{subfigure}[t]{0.5\linewidth}
        \centering
        \includegraphics[height=1.5in]{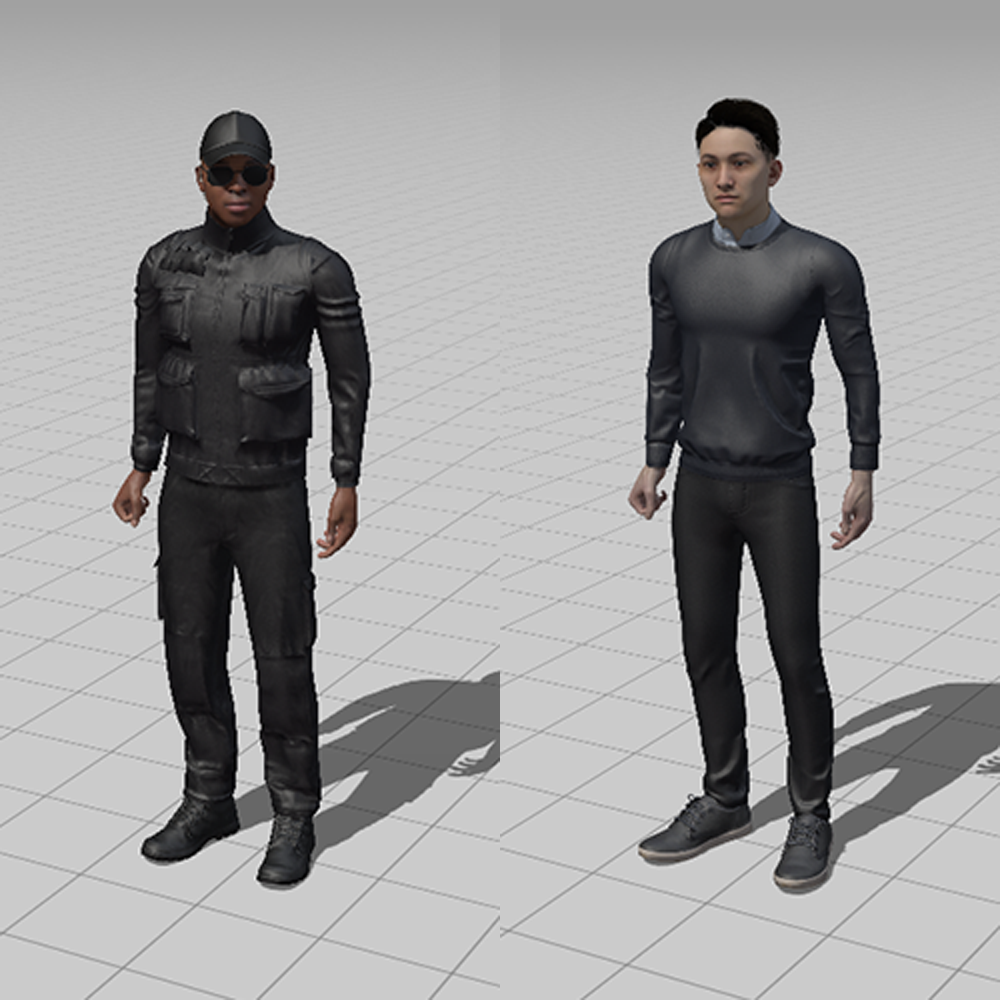}
        \caption{Partner and suspect NPC character models}
    \end{subfigure}%
    ~ 
    \begin{subfigure}[t]{0.5\linewidth}
        \centering
        \includegraphics[width=1\linewidth]{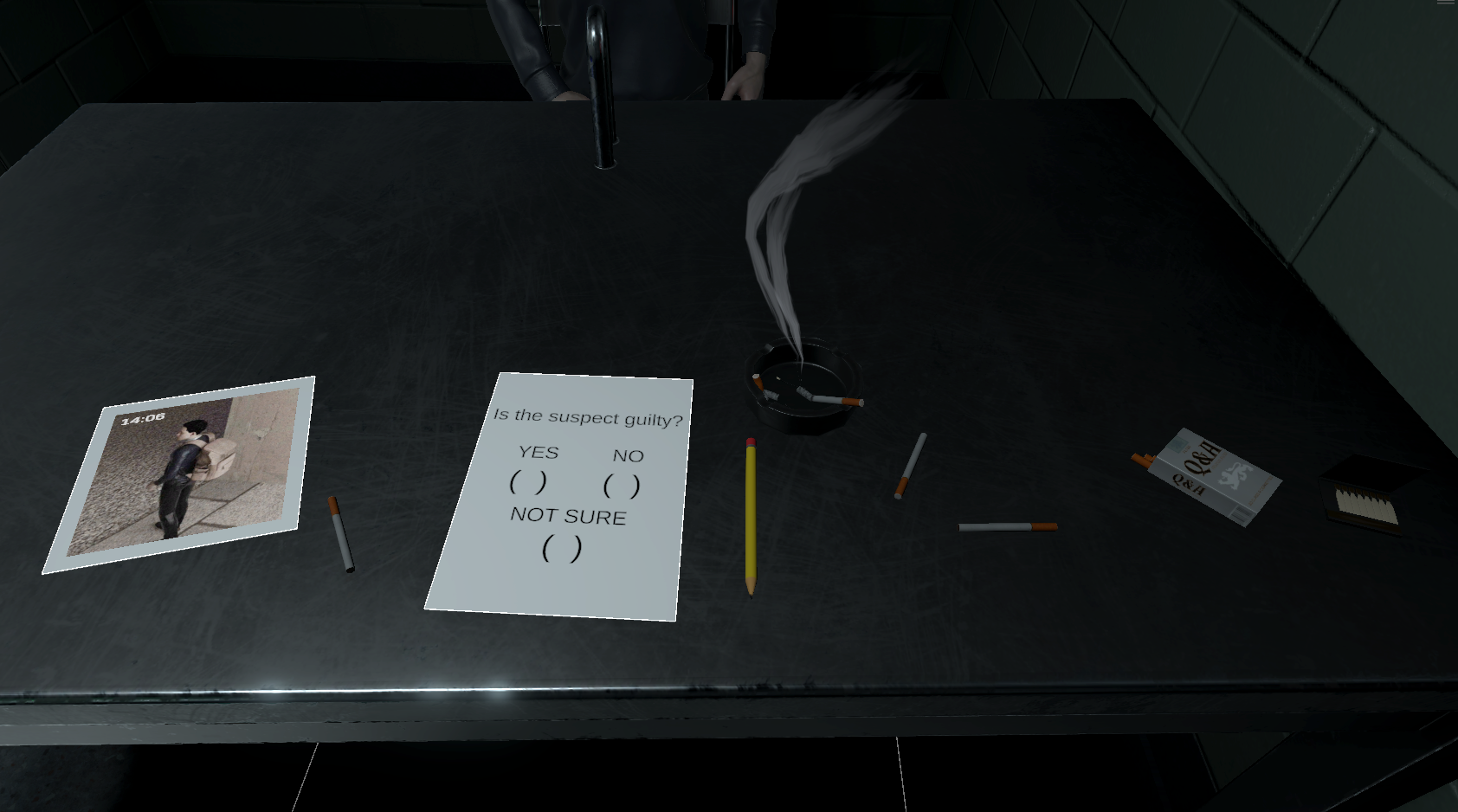}
        \caption{User's view (Photo of the crime, decision card, and "smokable cigarettes"}
    \end{subfigure}    
    
    \caption{The areas and characters featured in the simulator}
    
    \label{Rooms}
\end{figure*}

The application used in our study is a VR application depicting an interrogation situation between a police officer (study participant) and a crime suspect. The goal is to determine whether the suspect is guilty or innocent through interrogation. The application features 3D models of police interrogation and observation rooms and two NPC characters (Figure \ref{Rooms}.). The two NPCs can be interacted with and are controlled by AI (GPT-4 Turbo). The "suspect" NPC is the one the participant is interrogating while the "partner" NPC is helping the participant in the interrogation by providing additional information about the suspect and the crime, advice on what to ask next, reminding the participant about previous replies from the suspect, etc.

\subsection{The scenario}
The interrogation scenario evolves around a simple crime. The person (suspect NPC) being interrogated is suspected of stealing a case of beer from the local supermarket. Before the game is started, the researcher sets the suspect NPC to either guilty or innocent. The interrogator (participant) is not given any information beforehand on whether the suspect has committed this crime or not. The scenario starts with the player and the player's partner NPC in the observation room (Figure \ref{Rooms}.b). The NPC gives the player a quick description of the crime and why the suspect is there for questioning. The player can ask the partner NPC questions about the crime and the suspect's involvement. During this phase, the player is also taught how to question the suspect NPC (recording the question as audio) and how to contact the partner NPC for additional information and guidance. Once the participant is ready, they are moved to the interrogation room (Figure \ref{Rooms}.a) and the interrogation can begin. Now it is up to the participant to interrogate the suspect NPC with the help of the partner NPC to determine whether the suspect NPC is guilty or not.

\subsection{Implementation} \label{implementation}
The application was developed using Unity and the OpenXR plugin. It uses speech-to-text (STT), text-to-speech (TTS), and OpenAI GPT-4 Turbo \cite{openaiGPT4Turbo2024} that were implemented into Unity using the OpenAI API. STT is used to convert the user's questions into text format while TTS is used to convert GPT's text-form responses into speech for either of the NPCs. The application was built and tested using the Meta Quest 2 VR head-mounted display (HMD). The application's interrogation loop (starts when the participant enters the interrogation room) works as follows:

\begin{enumerate}
    \item The player's question is recorded as an audio clip
    \item The host encodes the audio clip and sends it to an LLM that converts it to text (STT)
    \item The LLM sends the generated text back to the host, and it is then sent to OpenAI GPT model.
    \item GPT interprets the question and formulates a response in text format.
    \item GPT model sends the response back to the host.
    \item The host sends the text format response to an LLM to be converted into audio (TTS)
    \item The LLM sends the generated audio file back to the host application.
    \item The suspect NPC answers the question by playing the audio file.
\end{enumerate}

\begin{figure}[H]
    \centering
    \includegraphics[width=0.9\linewidth]{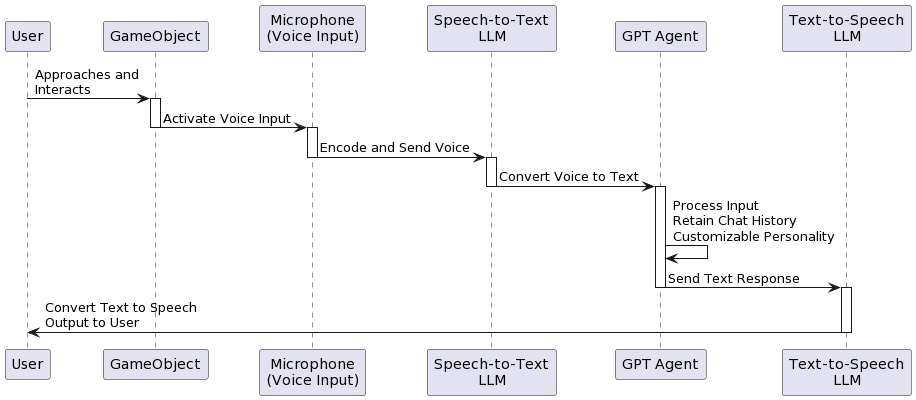}
    \caption{Sequence diagram of the interrogation loop}
    \label{sequence-diagram}
\end{figure}

A more detailed description of the interrogation loop can be observed in Figure \ref{sequence-diagram}.

At any point during the interrogation, the player has the option to consult the partner NPC for assistance. Interacting with the partner NPC follows the same process as outlined in the interrogation loop described above. The partner NPC is positioned in the observation room and has full access to the conversation logs, enabling them to hear and analyze everything discussed in the interrogation room. However, the suspect NPC cannot hear the exchanges between the player and the partner NPC, nor can they hear the partner NPC's responses to the player.


It is important to understand how the OpenAI GPT model maintains contextual awareness during the interrogation process. The implementation of this NPC "memory" is basically a text log of the conversation history (chat log). Each time the user interacts with the NPC, both the user's query and the NPC's response are appended to an array that stores the entire chat history. During subsequent interactions, instead of processing a single message, the entire chat history is sent to OpenAI GPT for processing. This mechanism enables the GPT model to retain context, giving the NPC the appearance of memory and situational awareness. However, due to the nature of this implementation, the processing time (latency) is expected to increase as the conversation progresses since the size of the chat history grows continuously over time.

\subsection{NPC personalities}

A key aspect of the application design was ensuring that the NPCs exhibit unique and distinguishable personalities. The initial idea had been to make the partner NPC Finnish; however, due to technical challenges with TTS and STT systems, it was decided that both the suspect and partner NPCs would communicate in English. To have some variety between the NPCs, the suspect NPC was trained to be an Australian exchange student but their personality would change according to the questions the interrogator asked each time the game was played. The suspect NPC was designed to show mysterious behavior, aiming to avoid detection and compelling the player to question their guilt or innocence. The character was initially developed using GPT-3.5 Turbo. However, this approach was later refined with GPT-4 prompts due to performance inconsistencies. Specifically, the initial implementation with GPT-3.5 Turbo resulted in behavior that was misaligned with design objectives; for instance, the NPC frequently used Australian slang to an extent that hindered comprehension. The simulator now employs GPT-4 Turbo to maintain the desired personality features and linguistic clarity.

Similarly to the suspect NPC, the partner NPC is designed with distinct personality characteristics, ensuring dynamic and varied interactions with the player. Their role involves offering strategic advice to assist the player in designing new approaches to extracting information from the suspect and advancing the interrogation process. In contrast to the suspect NPC, the partner NPC communicates in neutral, accent-free English to enhance clarity and accessibility. Like the suspect NPC, the partner NPC was trained using GPT-4 prompts.

\section{Method}
The user study was conducted in April 2024 in various locations. In each location, the testing procedure was standardized and identical to maintain the consistency and reliability of the results. Participants were tasked with determining the guilt or innocence of an AI-controlled NPC. However, the primary focus of this study was to evaluate the performance and believability of the AI NPCs rather than the participants' decision-making outcomes.
Participation in the study was open to individuals without specific restrictions, besides possessing sufficient English language proficiency to comprehend and communicate effectively. This requirement ensured that all participants could fully engage with the study materials and provide meaningful feedback on the AI NPCs' interactions.


\subsection{Participants}
A total of 18 participants (14 males and 4 females) aged between 20 and 25 years were recruited for this study. Recruitment was conducted at the University of Oulu and through the researchers' personal networks, which included colleagues and friends. This approach resulted in a convenience sample selected due to the project's strict time limitations. While this sampling method facilitated the timely acquisition of participants, it may limit the generalizability of the findings to broader populations.


\subsection{Study procedure}

The test session started with the participant reading and signing the content form. Next, the user would read through the instructions regarding the game. At this point, the participants could also ask questions if something was unclear to them. When they were ready, the game would be started. The participant would have up to 20 minutes to play the game to try and determine whether the suspect was either guilty or innocent. However, the game could end before the 20 minutes were up given that the participant had come to a conclusion regarding the guilt of the suspect. After finishing the game session, the participant would be asked to fill out three questionnaires: The System Usability Scale (SUS) \cite{SUS}, The Game Experience Questionnaire (GEQ, the social presence module) \cite{GameExperienceQuestionnaire}, and a Virtual Agent Believability Questionnaire \cite{BelievabilityOfVirtualAgents}. Following the questionnaires, we asked the participants to answer the following statements on a 5-point Likert scale (1: strongly disagree, 2: disagree, 3: neither agree nor disagree, 4: agree 5: strongly agree):
\begin{enumerate}    
    \item I felt that the response times negatively impacted realism.
    \item I would have enjoyed the game more if the response time were shorter.
    \item I felt the partner in the game was helpful in the interrogation.
\end{enumerate}

Additionally, we asked the users to evaluate how long on average did it take for the virtual agent(s) to respond. The response options were 1s, 5s, 10s, 15s, 20s, and more than 25s. They were also given the opportunity to leave feedback. Finally, the participants were rewarded a prize for participating in the study. Before the next experiment, the HMD along with the controllers are sanitized.

\section{Results}
The results gathered from the user tests are used to evaluate various aspects of the user experience (SUS and GEQ), the performance (latency data), and the believability of the virtual agents (the Virtual Agent Believability Questionnaire).

\subsection{Performance metrics} \label{performance_metrics}

We recorded separate latencies for STT, TTS, and the GPT model. From these latencies, we also calculated what we call "cycle latency". This cycle latency represents the total time the user has to wait to hear the response from the suspect to the question they've asked. So, in the interrogation loop, the cycle latency is the waiting time in between audio recording (interrogator's question) and playback (suspect's answer):  audio recording $\rightarrow \textbf{STT} \rightarrow \textbf{GPT} \rightarrow \textbf{TTS} \rightarrow$ audio playback. The full interrogation loop description can be found in Section \ref{implementation}. 

Naturally, all the measured latencies are affected by the length of the input (speech, text, generated reply), especially TTS.  Other contributing factors are the latency of the service and network latency.  The performance of OpenAI GPT is also affected by the cumulative size of the chat history. This is because in order for the model to have knowledge about prior questions and responses, all previous questions and GPT's responses (chat history) need to be sent along with the new question every time. Still, by averaging the total latencies, we can get a good overall picture of the performance of our system. The recorded latencies can be observed in Table \ref{table:Latencies}. For more context, latencies for STT, TTS, and OpenAI GPT plotted with their respective API call index can be found in Appendix (Figures \ref{fig:stt_latency_vs_index}, \ref{fig:tts_latency_vs_index}, and \ref{fig:gpt_latency_vs_index}.) as well as regression analysis for the TTS latency (Figure \ref{fig:GPTvsTTSlat}.) and a density plot comparing the distributions of the latencies (Appendix Figure \ref{fig:density}).

\begin{table}[htbp]
\centering
\scriptsize
\renewcommand{\arraystretch}{1.5}
\caption{Average, median, maximum, minimum, and standard deviation for the measured GPT, STT, TTS, and cycle latencies.}
\begin{tabular}{@{}c|ccccc@{}}
\toprule
\textbf{API type }& \textbf{Average} & \textbf{Median} & \textbf{Maximum }& \textbf{Minimum} & \textbf{Standard deviation }\\
 \midrule
GPT Model & 3113 ms & 2491 ms & 14918 ms & 696 ms & 2239 ms  \\
Speech-to-text & 1365 ms & 1290 ms & 3475 ms & 405 ms & 508 ms  \\
Text-to-speech & 2376 ms & 2087 ms & 12404 ms & 293 ms & 1216 ms \\
\midrule
Cycle latency & 6909 ms & 6066 ms & 24362 ms & 2021 ms & 3118 ms \\
\bottomrule

\end{tabular}
\label{table:Latencies}
\end{table}

It is important to mention that the game had a timeout mechanism that did not allow API calls to last longer than fifteen seconds. During user testing, this happened a few times with the GPT call in particular. As a result, the maximum latency of the GPT API is 14918 ms, close to the cutoff threshold of fifteen seconds, or 15000 ms. From Appendix Figure \ref{fig:api_boxplot}. We can also observe that the GPT API exhibited a large amount and variance in outliers compared especially to the STT latency.

\begin{table}[htbp]
\centering
\scriptsize
\renewcommand{\arraystretch}{1.3}
\caption{Total measured (cycle) latency, user evaluated latency, correct (closest) option, correct option - measured latency, user evaluated latency - measured latency, and error between user evaluated latency and correct option in relation to actual measured latency.}
\begin{tabular}{@{}c|P{1.1cm}P{1.5cm}P{1.0cm}P{2.2cm}P{2.2cm}P{2.2cm}@{}}
\toprule
\textbf{User}& \textbf{Measured latency (s)} & \textbf{User evaluated latency (s)} & \textbf{Correct option (s)}& \textbf{Measured latency - Correct option (s)} & \textbf{Measured latency - User evaluated latency (s)} & \textbf{Error (s)}\\
 \midrule
 
1 & 8.1 & 15 & 10 & $\lvert8.1-10\rvert\ $= 1.9 & $ \lvert8.1-15\rvert\ $= 6.9  & $ \lvert1.9-6.9\rvert\ $= 5.0  \\
2 & 5.9 & 10 & 5 & $ \lvert5.9-5 \rvert\ $= 0.9 & $ \lvert5.9-10\rvert\ $=  4.1  & $ \lvert0.9-4.1\rvert\ $= 3.2 \\
3 & 5.8 & 5 & 5 & $ \lvert5.8-5 \rvert\ $= 0.8 & $ \lvert5.8-5\rvert\ $= 0.8 & $ \lvert0.8-0.8\rvert\ $= 0  \\
4 & 8.8 & 10 & 10 & $ \lvert 8.8-10 \rvert\ $= 1.2 & $ \lvert8.8-10\rvert\ $=  1.2  & $ \lvert1.2-1.2\rvert\ $= 0 \\
5 & 7.7 & 1 & 10 & $ \lvert7.7-10 \rvert\ $= 2.3 & $ \lvert7.7-1\rvert\ $= 6.7 & $ \lvert2.3-6.7\rvert\ $= 4.4 \\
6 & 5.3 & 5 & 5 & $ \lvert5.3-5 \rvert\ $= 0.3 & $ \lvert5.3-5\rvert\ $= 0.3 & $ \lvert0.3-0.3\rvert\ $= 0\\
7 & 6.7 & 15 & 5 & $ \lvert6.7-5 \rvert\ $= 1.7 & $ \lvert6.7-15\rvert\ $= 8.3 & $ \lvert1.7-8.3\rvert\ $= 6.6\\
8 & 6.2 & 10 & 5 & $ \lvert6.2-5.0 \rvert\ $= 1.2 & $ \lvert6.2-10\rvert\ $= 3.8 & $ \lvert1.2-3.8\rvert\ $= 2.6\\
9 & 6.6 & 10 & 5 & $ \lvert6.6-5 \rvert\ $= 1.6 & $ \lvert6.6-10\rvert\ $= 3.4 & $ \lvert1.6-3.4\rvert\ $= 1.8\\
10 & 8.0 & 10 & 10 & $ \lvert8.0-10 \rvert\ $= 2.0 & $ \lvert8.0-10\rvert\ $=  2.0 & $ \lvert2.0-2.0\rvert\ $= 0\\
11 & 7.2 & 10 & 5 & $ \lvert7.2-5 \rvert\ $= 2.2  & $ \lvert7.2-10\rvert\ $= 2.8 & $ \lvert2.2-2.8\rvert\ $= 0.6\\
12 & 6.8 & 5 & 5 & $ \lvert6.8-5 \rvert\ $= 1.8 & $ \lvert6.8-5\rvert\ $= 1.8 & $ \lvert1.8-1.8\rvert\ $= 0\\
13 & 7.8 & 5 & 10 & $ \lvert7.8-10 \rvert\ $= 2.2 & $ \lvert7.8-5\rvert\ $= 2.8 & $ \lvert2.2-2.8\rvert\ $= 0.6\\
14 & 6.9 & 5 & 5 & $ \lvert6.9-5 \rvert\ $= 1.9 & $ \lvert6.9-5\rvert\ $= 1.9 & $ \lvert1.9-1.9\rvert\ $= 0\\
15 & 10.5 & 5 & 10 & $ \lvert10.5-10\rvert\ $= 0.5 & $ \lvert10.5-5\rvert\ $= 5.5 & $ \lvert0.5-5.5\rvert\ $= 5.0\\
16 & 7.6 & 15 & 10 & $ \lvert7.6-10\rvert\ $= 2.4 & $ \lvert7.6-15\rvert\ $= 7.4 & $ \lvert2.4-7.4\rvert\ $= 5.0\\
17 & 10.6 & 10 & 10 & $ \lvert10.6-10\rvert\ $= 0.6 & $ \lvert10.6-10\rvert\ $= 0.6 & $ \lvert0.6-0.6\rvert\ $= 0\\
18 & 8.2 & 10 & 10 & $ \lvert8.2-10\rvert\ $= 1.8 & $ \lvert8.2-10\rvert\ $= 1.8 & $ \lvert1.8-1.8\rvert\ $= 0\\ 
\bottomrule

\end{tabular}
\label{table:Evaluated latencies}
\end{table}

The users were also asked to evaluate how long, on average, the virtual agent(s) took to respond. The response options were 1s, 5s, 10s, 15s, 20s, and more than 25s. The average of the responses was \textbf{8.67} s and comparing this to the actual average latency of \textbf{6909} ms (Table \ref{table:Latencies}.) we can say that the users fairly accurately estimated the average latency. Per user average cycle latencies (measured latency) and user evaluated latency can be found in Table \ref{table:Evaluated latencies}.

As illustrated in Table \ref{table:Latencies}, 10 out of 18 participants provided incorrect latency evaluations. Among these, 7 participants overestimated the latency, while 3 underestimated it. Specifically, participant \#7 overestimated the latency by two response steps, and participant \#5 underestimated it by two response steps. The remaining 8 participants accurately selected the corresponding latency option. Considering the predefined response options were spaced 5 seconds apart, with an initial 4-second interval between 1 and 5 seconds, some degree of inaccuracy is anticipated. However, upon closer examination, 4 of 10 incorrect responses fall within an acceptable error margin. This acceptance is based on the proximity of the measured latencies to multiple response options and the inherent variability of latency during interactions. An acceptable error margin of 3.0 seconds was established, comprising 2.5 seconds for proximity to adjacent options and 0.5 seconds for user tolerance. For instance, participant \#13's evaluation of 5 seconds was 2.8 seconds away from the measured latency of 7.8 seconds, while the closest correct option (10 seconds) was 2.2 seconds away, resulting in an error of 0.6 seconds. With this error tolerance applied, the number of effectively correct evaluations increases to 12 out of 18 participants, demonstrating a higher accuracy level in user-perceived latency than initially observed. This suggests that users' perceptions of system latency are generally reliable despite the limitations imposed by the discrete response options and the inherent latency variability over time.

\subsection{Virtual Agent Believability Questionnaire}

This questionnaire measures various aspects that relate to the believability of virtual agents. The categories measured are. visual, behavior, awareness, social, intelligence, emotion, personality, agency, and overall believability. The questionnaire features 36 questions that are answered with a 7-point Likert scale ranging from "strongly disagree" to "strongly agree". The scores in the categories are summed up and then multiplied by a weight value and finally divided by the number or question in the category. The total believability is then calculated by summing up the category scores and dividing by the number of categories. The results can be found in Table \ref{table:Believability}.

The scoring of the original believability scale was not intuitive to us. For example, the minimum value of each category was not 0 so we decided to re-scale the results. Now, each sub-category and the total believability gives a score ranging from 0 to 10.

\begin{table}[htbp]
\caption{Virtual agent believability average scores for each group and for both groups. The maximum value for each category is 10.}
\centering
\scriptsize
\renewcommand{\arraystretch}{1.3}
\begin{tabular}{@{}c|cccc@{}}
\toprule
\textbf{Category} &\textbf{ Guilty suspect group} & \textbf{Innocent suspect group} & \textbf{Both groups}\\
 \midrule
Visual properties & 6.98 & 5.99 & 6.48  \tabularnewline
Behavior & 8.11 & 8.07 & 8.09 \tabularnewline
Awareness & 5.44 & 7.11 & 6.28  \tabularnewline
Social relationships & 8.43 & 8.06 & 8.24 \tabularnewline
Intelligence & 8.15 & 7.84 & 7.99  \tabularnewline
Emotion & 6.34 & 5.97 & 6.16 \tabularnewline
Personality & 5.34 & 6.14 & 5.74  \tabularnewline
Agency & 5.37 & 5.31 & 5.34 \tabularnewline
Overall believability & 7.96 & 7.41 & 7.69  \tabularnewline
\midrule
Total believability & 6.62 & 6.72 & 6.67  \tabularnewline
\bottomrule

\end{tabular}
\label{table:Believability}
\end{table}

\subsection{System Usability Scale}

The SUS questionnaire features 10 questions and uses a 5-point Likert scale. It is used to measure how the user perceives different aspects of the system's usability such as effectiveness, efficiency, and satisfaction. With this questionnaire, we wanted to study how the participants felt about the communication systems for interrogation and talking with the partner NPC as well as how the latencies (TTS, STT, GPT) affected the usability. The user scores ranged from \textbf{47.50} to \textbf{97.50}, the average being \textbf{79.44}. This rates our application as "good" (grade B) and just short of excellent (> 80.3, grade A).

\subsection{Game Experience Questionnaire} \label{GEQ}

From GEQ, we used the social presence module. With this module, we could investigate the psychological and behavioral involvement of the user with other social entities, in our case, the AI-controlled partner and suspect NPCs. The participants scored (averages) the categories as follows: 
\begin{itemize}
    \item Pyschologial involvement - Empathy: \textbf{1.23}
    \item Pyschologial involvement - Negative feelings: \textbf{1.26}
    \item Behavioral involvement: \textbf{3.03}
\end{itemize}
The maximum score for each category is 4.00. We can see that the psychological involvement categories had low scores (1.23 and 1.26) while behavioral involvement was relatively high at 3.03. 

\subsection{Additional questions} \label{Additional_Q}

With the additional questions, we wanted the users to evaluate a few specific issues. The participants were asked to rate the following statements on a 5-point Likert scale (1: strongly disagree, 2: disagree, 3: neither agree nor disagree, 4: agree 5: strongly agree):
\begin{itemize}
    \item I felt that the response times negatively impacted realism: \textbf{3.33}
    \item I would have enjoyed the game more if the response time were shorter: \textbf{3.61}
    \item I felt the partner in the game was helpful in the interrogation: \textbf{4.28}    
\end{itemize}

At the end of the interrogation, the participants had to make a decision about the guilt/innocence of the suspect NPC. The decision accuracy of the groups was as follows:

\begin{itemize}
    \item Suspect \textbf{GUILTY}: Correct: \textbf{66.7}\%, Incorrect: \textbf{22.2}\%, Not sure: \textbf{11.1}\%
    \item Suspect \textbf{INNOCENT}: Correct: \textbf{44.4}\%, Incorrect: \textbf{44.4}\%, Not sure: \textbf{11.1}\%   
\end{itemize}

\section{Discussion}

\subsection{Latencies}

From Table \ref{table:Latencies}. and Figure \ref{fig:density}. (Appendix), we can observe that in our application the STT latency was the lowest by far, followed by TTS and GPT latency being the highest on average. The standard deviation in latency also follows the same pattern with STT having quite a low deviation with \textbf{508} ms, TTS with \textbf{1216} ms, and GPT model having the highest deviation with \textbf{2239} ms. The high standard deviation in GPT latency can be attributed to the variability in the amount of data (chat history) processed to generate responses to the interrogator’s questions.  Specifically, as the duration of the interrogation increases, the chat history grows larger, leading to progressively higher latency for OpenAI GPT. This behavior is clearly seen in Figure \ref{fig:GPTvsINDEX}. (Appendix) where the GPT model latency is trending upward over time.

The TTS latency (Appendix Figure \ref{fig:tts_latency_vs_index}.) shows a similar but not as steep a trend as GPT model. To explore why they both exhibited this similar trend as the game progresses we performed a regression analysis to analyze the relationship between the length of the GPT response and the TTS latency. The analysis was performed using the Scikit-learn \cite{Scikit-learn} Python library. The resulting regression coefficient was \textbf{10.23} with an intercept of \textbf{581.90} and an R-squared value of \textbf{0.75}. These results indicate a strong positive correlation between the length of the GPT response and the TTS latency (Appendix Figure \ref{fig:GPTvsTTSlat}). Naturally, we expect the correlation between the length of the response and TTS latency to exist simply because the longer the response, the longer it will take for it to be converted into speech. But, this does not explain why the TTS latency increases over time. Because the TTS latency is mostly affected by the GPT response length we decided to investigate whether the GPT response length also increased over time. In Figure \ref{fig:GPTvsINDEX}. (Appendix) we can see that the response length is indeed increasing (upward trend) as the GPT index (index indicates the number of the request) increases.  

While all these latencies are interesting, perhaps the most important one is the cycle latency which is the summation of all these latencies and, in essence, the waiting time for the user between their question to the suspect and the suspect's answer. All latencies are represented in Table \ref{table:Latencies}. and from there we can see that the average waiting time (cycle latency average) was \textbf{6909} ms which, if it were a waiting time in a conversation between two real people, we would already consider a bit long. However, the minimum waiting time of \textbf{2021} ms is, in our opinion, an excellent value. And just like in a real conversation, there is a "natural" variation in the cycle latency that can be seen from the standard deviation of \textbf{3118} ms. However, from the perspective of the user, 3 seconds can feel like a big difference especially if they are expecting consistent response times. Unfortunately, we also cannot escape the fact that the maximum cycle latency was very high at over \textbf{24} s. High latency and fluctuations in the response times can be frustrating to the user and can make the NPC feel more like a machine instead of a real person. The fluctuations most likely caused some participants to inaccurately estimate the average cycle latency they experienced during their interrogation and this can be seen in Section \ref{performance_metrics} and Table \ref{table:Evaluated latencies}. Still, 66.7\% (12/18) of participants were able to, within acceptable error limits, accurately estimate the average cycle latency they experienced throughout the interrogation.

\subsection{NPC believability}

Along with the NPC performance, its believability was an important part of this study. We measured the believability of our AI NPCs using the Virtual Agent Believability Questionnaire (Table \ref{table:Believability}). The areas we would characterize as good (score \textgreater \textbf{7.5}) were behavior, social relationships, intelligence, and overall believability. The areas in which our NPCs performed the worst were personality and agency, although with scores of \textbf{5.74} and \textbf{5.34} respectively, they would still rank as average. Visual properties (\textbf{6.48}), awareness (\textbf{6.28}), and emotion (\textbf{6.16)} also rank as average.

If we look at the traits that can make an AI seem more human, like emotion and personality, the scores (\textbf{6.16} and \textbf{5.74}) do not accurately represent what was achieved with our AI training. For personality, the participants answered that the NPCs didn't feel sympathetic, warm, extroverted, or enthusiastic. But this is exactly how the AIs were trained to act so despite the low score, we actually achieved our goal of making the partner NPC act complacent and taciturn while the suspect NPC was trained to be nervous and socially awkward. For emotion, the answers were more neutral in terms of the NPCs being able to have and express emotions that were understandable and appropriate to the situation. These results also make sense given the AI training, especially since the suspect was meant to be socially awkward and the partner reserved and "uncommunicative". 

We did not see large differences in the believability categories between the two groups (suspect guilty/innocent). The biggest differences between the two groups were in the areas of visual properties (guilty: \textbf{6.98}, innocent: \textbf{5.99}) and awareness (guilty: \textbf{5.44}, innocent: \textbf{7.11}). The reason why the guilty group rated visual properties to be somewhat better than the innocent group is not apparent, since the visual properties of the NPCs were exactly the same between the versions. Perhaps the difference can be attributed to the relatively low number of participants and the low number of questions in that section. Both of these aspects introduce a higher possible variance. As for awareness, there is a possibility that participants interpreted the questions differently. For example, for a question that asks if the virtual agent was aware of the presence of other virtual agents the participants' answers were very divided. If the participants only considered the suspect, their answer would most certainly be negative, but if they based their answer on the partner they would most likely feel that the NPC was aware of the suspect.

Overall we consider the believability of the NPCs was fairly good. But it could certainly be improved upon especially in the areas of agency and awareness, since our NPCs were completely stationary and did not take any actions or movements independently. Additionally, all the actions the NPCs took were related to speech. It is also an important aspect to consider regarding the results, that they were achieved by almost completely focusing on the speech and cognitive abilities of the NPCs, and everything else was secondary or non-existent in the development. It is also worth noting that some of the questions included in the Virtual Agent Believability questionnaire and other questionnaires were not appropriate for our game, but we included them nevertheless to ensure the integrity of the questionnaires

\subsection{Usability and game experience}

The SUS score for our application was\textbf{ 79.44}, indicating good usability. Players found the application easy to learn, the mechanics easy to understand, and felt satisfied using the system. Good usability is particularly noteworthy because many of the test participants had little to no prior experience with VR. For example, a significant aspect of usability for VR applications is the controls. In VR, the buttons must be especially intuitive since players (usually) cannot see the controllers while wearing the HMD. Still, there were a few instances where test participants needed assistance with the buttons.


The scores relating to social presence (game experience)can be found in Section \ref{GEQ}. Participants perceived high behavioral involvement with the NPCs, the score being \textbf{3.03} out of 4.00 for both groups combined. There was no significant difference between the groups. Participants did not experience much negative feelings towards the NPCs, scoring \textbf{1.26} out of 4.00. This can be considered a good result since it was not our purpose to make the NPCs unlikable, but rather neutral. To demonstrate, had the crime chosen for the game been more severe, it would have been preferable if the participants felt negative feelings towards the suspect. However, we did not want to choose a more severe crime than shoplifting since it might not have been appropriate for the context, and it would have been distasteful to have the test participant play a potentially disturbing game, especially since that was not the goal or focus of the study. Interestingly, the innocent group had more negative feelings towards the NPCs than the guilty group. This could indicate that the negative feelings felt towards the NPCs were not out of dislike for their actions, but rather of frustration towards them, produced by not being able to get the suspect to confess.

The score for empathy was similar to negative feelings, meaning that participants did not feel much empathy towards the NPCs. The innocent group felt a bit more empathy compared to the guilty group, which is understandable since they were more likely to view the suspect as innocent, as can be seen from their final decisions regarding the guilt of the suspect (Section \ref{Additional_Q}). There was one question in the empathy section that did not fit well into the context of our game. This question asks if the player felt that they admired the NPCs; this produced heavily negative answers, possibly skewing the score for empathy. We did not intend any of the NPCs to produce feelings of admiration in the player, but it is interesting to note, that there were three participants that felt some level of admiration towards the NPCs.

\subsection{Additional questions and user feedback}

As we could already see from the latency measurements, the waiting time could often be quite long. The effect of this can be observed from how the participants answered the questions about the latency's effect on realism (3.33/5) and if they would have enjoyed the game more if the latency was lower (3.61/5). It is quite clear that in order for an AI NPC like ours to seem more human-like, the latency of the system needs to be lower than what we were able to achieve. As for our partner NPC, the results were very encouraging, as the participants felt like it was very helpful in the interrogation (4.28/5).

\subsection{Future of AI NPCs}

According to our results regarding both latencies and believability, the future of AI NPCs seems very interesting. Having full conversations with a human-feeling AI NPC might not yet be possible, at least for the layman with no access to computers with advanced computing power, but for regular day-to-day chatting, this should already be possible. The latencies are already low enough to enable human-like quick responses from the NPCs, as long as the question/response does not require a large amount of context to be processed. As can be seen from Table \ref{table:Evaluated latencies}. the average latencies for the players seem somewhat high, but we need to bear in mind that these are average values for the complete interrogation, and the latency is fairly low in the beginning and as the interrogation goes on, the latency can get very high when there is a lot of context processing involved with each new question. 

For immersive applications that require the user to observe and process information, NPCs like our interrogation partner, could be very valuable. These NPCs could assist the user, for example, by answering the user's questions about the data they are seeing, giving the user advice on how to proceed, reminding the user of tasks that they need to perform based on certain information, etc. This type of AI assistant would help in improving the user's situational awareness which is maybe not that important in leisurely games but is an extremely important factor in serious applications. These types of actions should already be possible to implement to work very fast and if the AI does not need to feel like a human, the processing speed would be even faster.

\subsection{Limitations}

This user study used a prototype version of an "interrogation simulator" that focused on the player's interaction with AI NPCs. Thus, some aspects like aesthetics, NPC movement, etc. were not focused on and the believability results concerning these aspects should be taken with a grain of salt. Additionally, the sample size in our study was fairly small which prevents us from drawing concrete conclusions about the believability of our NPCs. 

\section{Conclusions}
This study developed and evaluated a VR interrogation simulator featuring AI-powered NPCs, focusing on their believability and system performance. The primary objectives were determining whether GPT-controlled NPCs could emulate human-like behavior and responses and whether the system's latency was low enough to maintain the illusion of real-time interaction.
In terms of system performance, the average latency during the interrogation was approximately 7 seconds, which, while somewhat high, is mitigated by a minimum latency of around 2 seconds and a standard deviation of 3 seconds. These metrics suggest that the system can achieve human-like responsiveness under optimal conditions, though the increasing latency over extended interactions remains a challenge. Despite this limitation, the performance results indicate a promising foundation for creating responsive and natural NPC interactions.\\
Believability was another critical metric, with the NPCs achieving an overall score of 6.67/10. Particularly strong ratings were observed in categories like social relationships (8.24/10), behavior (8.09/10), and intelligence (7.99/10), suggesting that participants thought the NPCs came across as more than AI and they perceived as behaving and socializing in human-like ways. Lower scores in emotion (6.16/10) and personality (5.74/10) align with the NPCs' intentional design to remain neutral and unemotional, supporting the idea that these dimensions were not primary goals for this study.
Overall, the results demonstrate the potential of GPT-powered AI NPCs to achieve a high degree of realism and interaction in VR environments while highlighting areas for improvement, particularly in latency management and enhancing emotional and personality-driven interactions. These findings provide valuable insights for the continued development of AI-driven NPCs capable of delivering immersive and believable user experiences.

\section*{Acknowledgments}

This work has been partially funded by the European Commission grant: A Cognitive Detection System for Cybersecure Operational Technologies (IDUNN) (101021911), funded by the European Commission. The authors declare no potential conflicts of interest with respect to the research, authorship, and/or publication of this article.

\newpage
\appendix
\section{}
\label{app1}
\begin{figure}[H]
  \begin{center}
    \includegraphics[width=14cm]{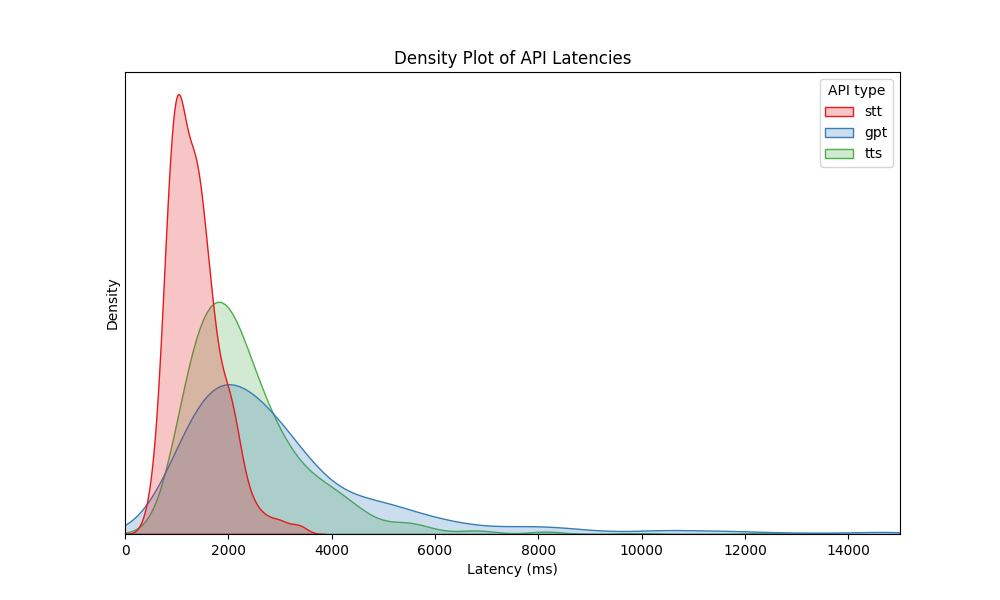}
  \end{center}
    \caption[]{Density plot of the duration of the different API calls (STT in red, GPT in blue, and TTS in green).}
    \label{fig:density}
\end{figure}
\vspace{-5mm}

\begin{figure}[H]
  \begin{center}
    \includegraphics[width=14cm]{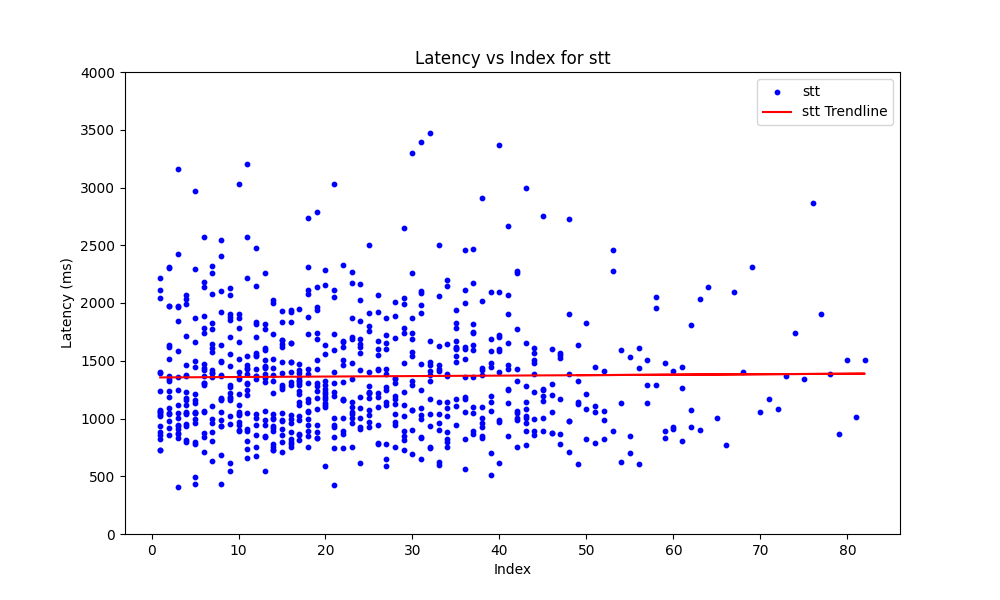}
  \end{center}
    \caption[]{Latency of STT call vs index.}
    \label{fig:stt_latency_vs_index}
\end{figure}

\begin{figure}[H]
  \begin{center}
    \includegraphics[width=14cm]{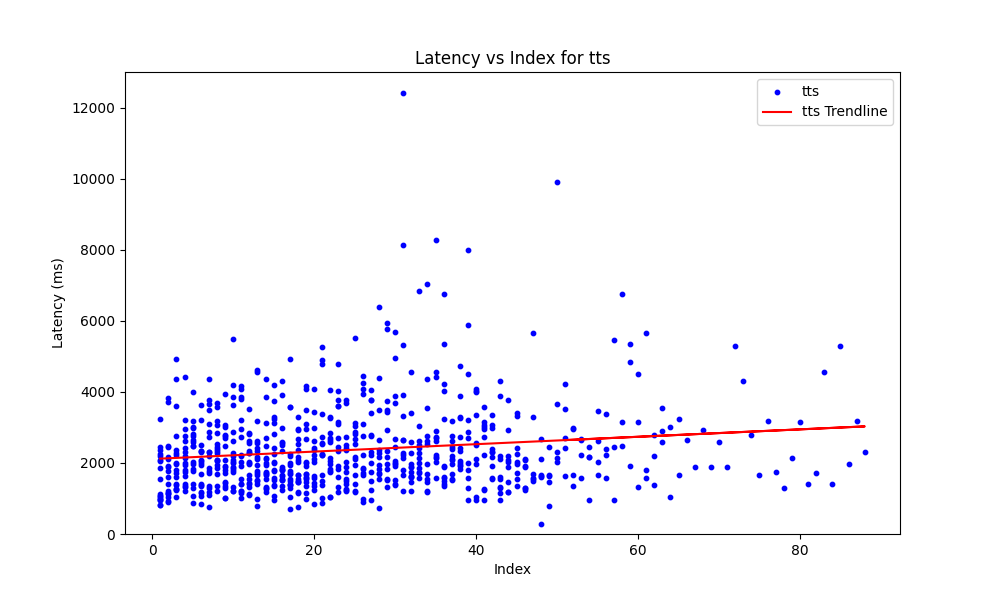}
  \end{center}
    \caption[]{Latency of TTS call vs index.}
    \label{fig:tts_latency_vs_index}
\end{figure}

\begin{figure}[H]
  \begin{center}
    \includegraphics[width=14cm]{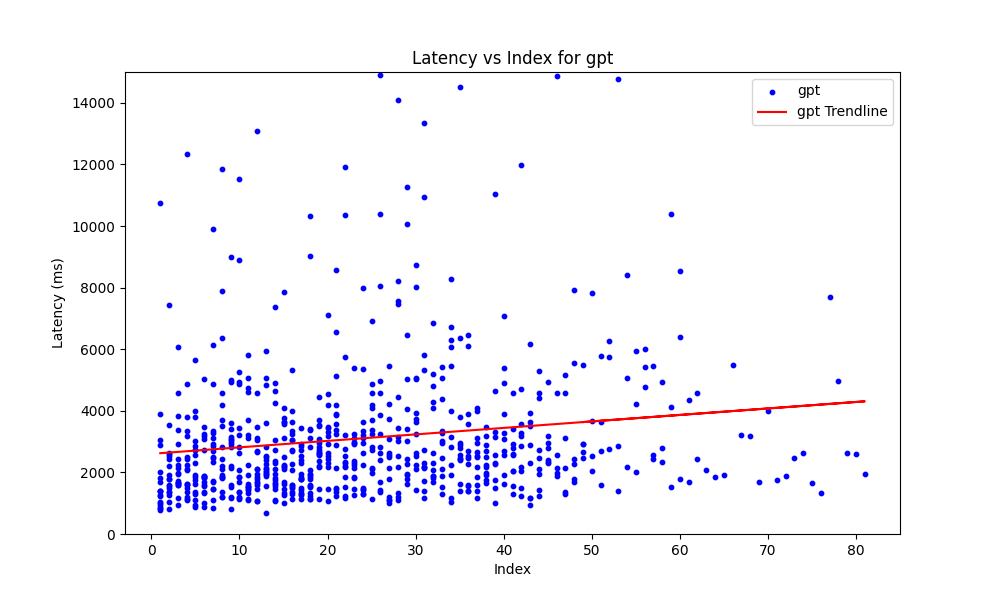}
  \end{center}
    \caption[]{Latency of GPT call vs index.}
    \label{fig:gpt_latency_vs_index}
\end{figure}

\vspace{-5mm}
\begin{figure}[H]
  \begin{center}
    \includegraphics[width=14cm]{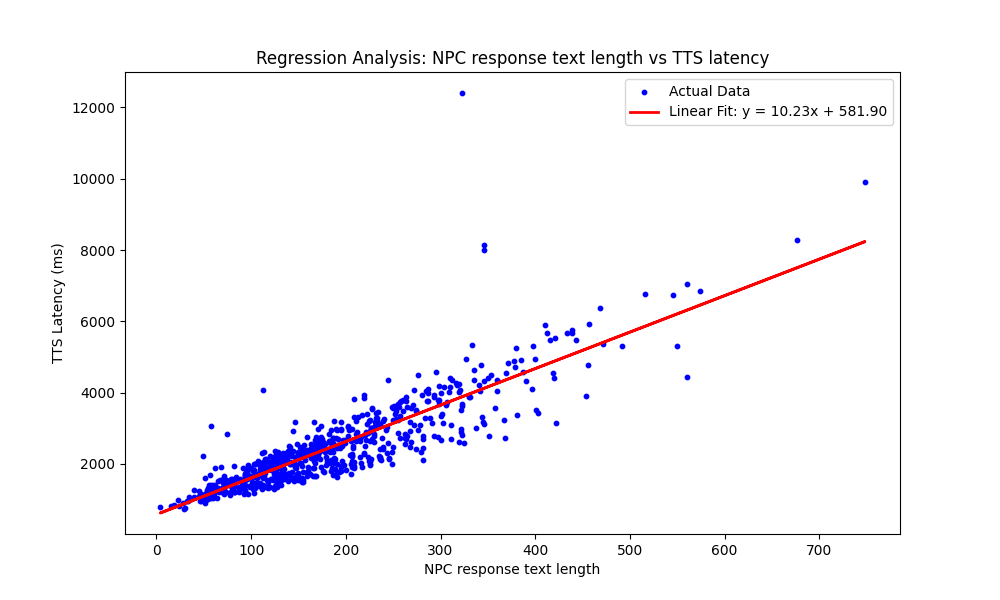}
  \end{center}
    \caption[]{Regression analysis of GPT response length vs. TTS latency.}
    \label{fig:GPTvsTTSlat}
\end{figure}

\begin{figure}[H]
  \begin{center}
    \includegraphics[width=14cm]{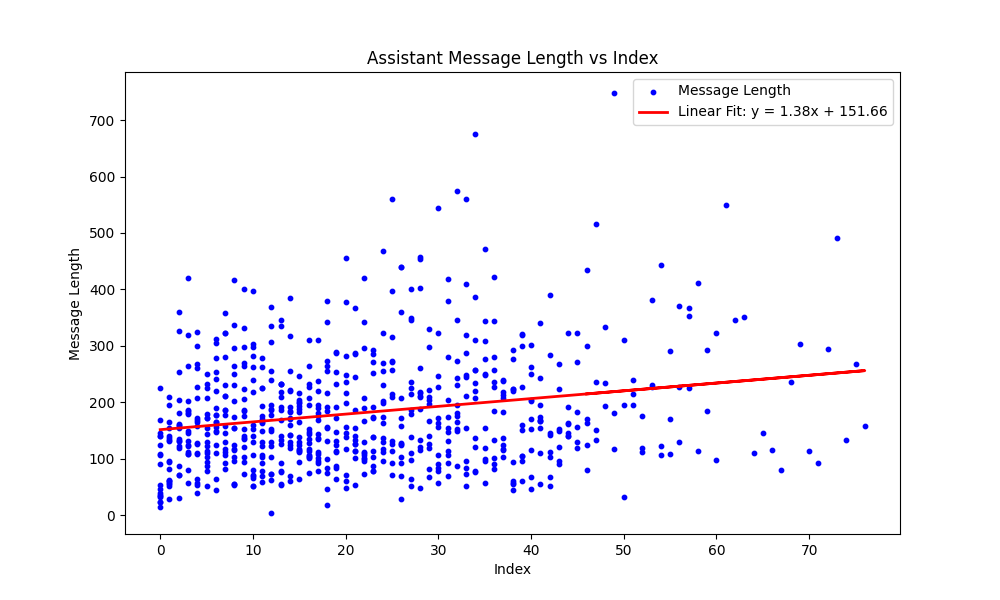}
  \end{center}
    \caption[]{GPT response length vs index in the game.}
    \label{fig:GPTvsINDEX}
\end{figure}

\begin{figure}[H]
  \begin{center}
    \includegraphics[width=14cm]{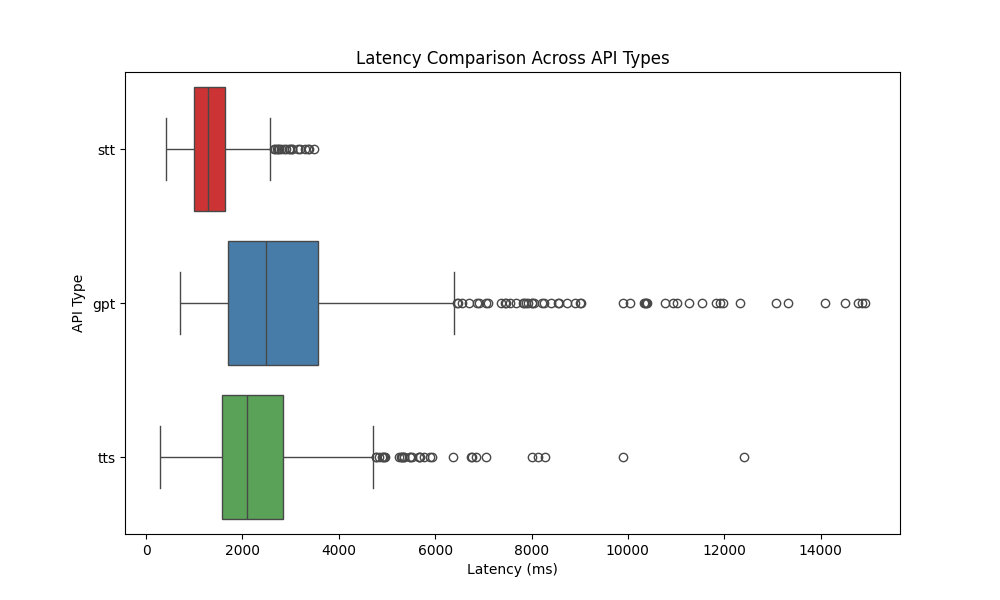}
  \end{center}
    \caption[]{Box plot of the duration of the different API calls.}
    \label{fig:api_boxplot}
\end{figure}



\newpage
\bibliographystyle{elsarticle-num} 
\bibliography{sn-bibliography}

\begin{thebibliography}{10}
\expandafter\ifx\csname url\endcsname\relax
  \def\url#1{\texttt{#1}}\fi
\expandafter\ifx\csname urlprefix\endcsname\relax\def\urlprefix{URL }\fi
\expandafter\ifx\csname href\endcsname\relax
  \def\href#1#2{#2} \def\path#1{#1}\fi

\bibitem{lapeyrade2023non}
S.~Lapeyrade, C.~Rey, Non-player character decision-making with prolog and ontologies, in: 2023 IEEE Conference on Games (CoG), IEEE, 2023, pp. 1--2.
\newblock \href {https://doi.org/https://doi.org/10.1109/CoG57401.2023.10333221} {\path{doi:https://doi.org/10.1109/CoG57401.2023.10333221}}.

\bibitem{karaca2023ai}
Y.~Karaca, D.~Derias, G.~Sarsar, Ai-powered procedural content generation: Enhancing npc behaviour for an immersive gaming experience, Available at SSRN 4663382 (2023).
\newblock \href {https://doi.org/https://dx.doi.org/10.2139/ssrn.4663382} {\path{doi:https://dx.doi.org/10.2139/ssrn.4663382}}.

\bibitem{filipovic2023role}
A.~Filipovi{\'c}, \href{https://www.ceeol.com/search/article-detail?id=1201751}{The role of artificial intelligence in video game development}, Kultura Polisa 20~(3) (2023) 50--67.
\newline\urlprefix\url{https://www.ceeol.com/search/article-detail?id=1201751}

\bibitem{zeng2023review}
G.~Zeng, A review of ai-based game npcs research, Applied and Computational Engineering 15 (2023) 155--159.
\newblock \href {https://doi.org/https://doi.org/10.54254/2755-2721/15/20230827} {\path{doi:https://doi.org/10.54254/2755-2721/15/20230827}}.

\bibitem{xu2023exploring}
W.~Xu, K.~Yu, X.~Meng, D.~Monteiro, D.~Kao, H.-N. Liang, Exploring the effect of the group size and feedback of non-player character spectators in virtual reality exergames, Frontiers in Psychology 14 (2023) 1079132.
\newblock \href {https://doi.org/https://doi.org/10.3389/fpsyg.2023.1079132} {\path{doi:https://doi.org/10.3389/fpsyg.2023.1079132}}.

\bibitem{guo2023s}
Z.~Guo, W.~Xu, J.~Zhang, H.~Wang, C.-H. Lo, H.-N. Liang, Who’s watching me?: Exploring the impact of audience familiarity on player performance, experience, and exertion in virtual reality exergames, in: 2023 IEEE International Symposium on Mixed and Augmented Reality (ISMAR), IEEE, 2023, pp. 622--631.
\newblock \href {https://doi.org/https://doi.org/10.1109/ISMAR59233.2023.00077} {\path{doi:https://doi.org/10.1109/ISMAR59233.2023.00077}}.

\bibitem{moon2018reviews}
J.~Moon, Reviews of social embodiment for design of non-player characters in virtual reality-based social skill training for autistic children, Multimodal Technologies and Interaction 2~(3) (2018) 53.
\newblock \href {https://doi.org/https://doi.org/10.3390/mti2030053} {\path{doi:https://doi.org/10.3390/mti2030053}}.

\bibitem{riyan2023implementation}
T.~S. Riyan, A.~Pardede, F.~Y. Manik, Implementation of finite state machine models on the artificial intelligence system of characters in the game" mmorpg" using rpg maker, Journal of Artificial Intelligence and Engineering Applications (JAIEA) 3~(1) (2023) 287--291.
\newblock \href {https://doi.org/https://doi.org/10.59934/jaiea.v3i1.311} {\path{doi:https://doi.org/10.59934/jaiea.v3i1.311}}.

\bibitem{ribeiro2023virtual}
T.~Ribeiro~de Oliveira, B.~Biancardi~Rodrigues, M.~Moura~da Silva, R.~Antonio N.~Spinass{\'e}, G.~Giesen~Ludke, M.~Ruy Soares~Gaudio, G.~Iglesias Rocha~Gomes, L.~Guio~Cotini, D.~da~Silva~Vargens, M.~Queiroz~Schimidt, et~al., Virtual reality solutions employing artificial intelligence methods: A systematic literature review, ACM Computing Surveys 55~(10) (2023) 1--29.
\newblock \href {https://doi.org/https://doi.org/10.1145/3565020} {\path{doi:https://doi.org/10.1145/3565020}}.

\bibitem{zheng2024memoryrepository}
S.~Zheng, K.~He, L.~Yang, J.~Xiong, Memoryrepository for ai npc, IEEE Access (2024).
\newblock \href {https://doi.org/https://doi.org/10.1109/ACCESS.2024.3393485} {\path{doi:https://doi.org/10.1109/ACCESS.2024.3393485}}.

\bibitem{ong2022dragonfly}
H.~S. Ong, H.~Junaedi, J.~Santoso, Dragonfly algorithm for crowd npc movement simulation in metaverse, Bulletin of Social Informatics Theory and Application 6~(1) (2022) 76--83.
\newblock \href {https://doi.org/https://doi.org/10.31763/businta.v6i1.551} {\path{doi:https://doi.org/10.31763/businta.v6i1.551}}.

\bibitem{strojny2020moderators}
P.~M. Strojny, N.~Du{\.z}ma{\'n}ska-Misiarczyk, N.~Lipp, A.~Strojny, Moderators of social facilitation effect in virtual reality: Co-presence and realism of virtual agents, Frontiers in psychology 11 (2020) 1252.
\newblock \href {https://doi.org/https://doi.org/10.3389/fpsyg.2020.01252} {\path{doi:https://doi.org/10.3389/fpsyg.2020.01252}}.

\bibitem{kim2020navigating}
J.-H. Kim, J.~Lee, S.-J. Kim, Navigating non-playable characters based on user trajectories with accumulation map and path similarity, Symmetry 12~(10) (2020) 1592.
\newblock \href {https://doi.org/https://doi.org/10.3390/sym12101592} {\path{doi:https://doi.org/10.3390/sym12101592}}.

\bibitem{breves2020reducing}
P.~Breves, Reducing outgroup bias through intergroup contact with non-playable video game characters in vr, Presence 27~(3) (2020) 257--273.
\newblock \href {https://doi.org/https://doi.org/10.1162/pres\_a\_00330} {\path{doi:https://doi.org/10.1162/pres\_a\_00330}}.

\bibitem{oumaima2023application}
D.~Oumaima, L.~Mohamed, H.~Hamid, H.~Mohamed, Application of artificial intelligence in virtual reality, in: International Conference on Trends in Sustainable Computing and Machine Intelligence, Springer, 2023, pp. 67--85.
\newblock \href {https://doi.org/https://doi.org/10.1007/978-981-99-9436-6\_6} {\path{doi:https://doi.org/10.1007/978-981-99-9436-6\_6}}.

\bibitem{pashentsev2023metaverses}
E.~Pashentsev, Metaverses, artificial intelligence and challenges to psychological security, Politika nacionalne bezbednosti 25~(2) (2023).
\newblock \href {https://doi.org/https://doi.org/10.5937/pnb25-46760} {\path{doi:https://doi.org/10.5937/pnb25-46760}}.

\bibitem{openaiGPT4Turbo2024}
OpenAI, \href{https://openai.com/}{Gpt-4 turbo}, large language model (2024).
\newline\urlprefix\url{https://openai.com/}

\bibitem{SUS}
J.~Brooke, Sus: A quick and dirty usability scale, Usability evaluation in industry (1996).
\newblock \href {https://doi.org/https://rickvanderzwet.nl/trac/personal/export/104/liacs/hci/docs/SUS-questionaire.pdf} {\path{doi:https://rickvanderzwet.nl/trac/personal/export/104/liacs/hci/docs/SUS-questionaire.pdf}}.

\bibitem{GameExperienceQuestionnaire}
W.~A. IJsselsteijn, Y.~A.~W. de~Kort, K.~Poels, The Game Experience Questionnaire (2013).
\newblock \href {https://doi.org/https://research.tue.nl/en/publications/the-game-experience-questionnaire} {\path{doi:https://research.tue.nl/en/publications/the-game-experience-questionnaire}}.

\bibitem{BelievabilityOfVirtualAgents}
S.~Guo, N.~Adamo, C.~Mousas, Developing a scale for measuring the believability of virtual agents, in: J.-M. Normand, M.~Sugimoto, V.~Sundstedt (Eds.), International Conference on Artificial Reality and Telexistence, Eurographics Symposium on Virtual Environments, Purdue University, 2023, p.~52.

\bibitem{Scikit-learn}
Scikit-learn, {Scikit-learn} web page, URL: \url{https://scikit-learn.org/stable/}, accessed: 16.6.2024.

\end{thebibliography}






\end{document}